\documentclass[12pt]{JHEP3}

\usepackage[normalem]{ulem}
\usepackage{amsmath}
\usepackage{amsthm}
\usepackage{amssymb}
\usepackage{enumerate}
\usepackage{bm,longtable,enumerate}
\usepackage{amsmath,amssymb}

\newcommand{\beqa}{\begin{eqnarray}}
\newcommand{\eeqa}{\end{eqnarray}}

\newcommand\tr{\mathop{\mathrm{Tr}}}

\newcommand{\beq}{\begin{equation}}
\newcommand{\eeq}{\end{equation}}

\def\ss{\scriptscriptstyle}

\title{DBI equations and holographic DC conductivity}

\author{Alfonso Ballon-Bayona  \\
ICTP South American Institute for Fundamental Research, \\
Instituto de F\'isica Te\'orica, Universidade Estadual Paulista, \\
01140-070 S˜ao Paulo, SP, Brazil.

\email{aballonb@ift.unesp.br}  }

\author{Cristine N. Ferreira$^{1,2}$\\
$^1$N\'ucleo de Estudos em F\'isica, Instituto Federal de Educa\c{c}\~{a}o,
Ci\^encia e Tecnologia Fluminense, \\ Rua Dr. Siqueira, 273, Campos dos
Goytacazes, \\
Rio de Janeiro, Brazil, CEP 28030-130 \\
$^2$The Abdus Salam
International Centre for Theoretical Physics\\
Strada Costiera, 11
I - 34151 Trieste Italy

\email{cferreir@ictp.it} }

\author{Victor J.  Vasquez Otoya \\
Instituto Federal de Educa\c{c}\~{a}o, Ci\^encia e Tecnologia do Sudeste de Minas Gerais, \\
Rua Bernardo Mascarenhas 1283, 36080-001, \\
Juiz de Fora, MG, Brasil

\email{victor.vasquez@ifsudestemg.edu.br} }

\abstract{We provide a simple method for writing the Dirac-Born-Infeld (DBI) equations of a Dp-brane in an arbitrary static background whose metric depends only on the holographic radial coordinate $z$.
Using this method we revisit the Karch-O'Bannon's procedure to calculate the DC conductivity in the presence of constant electric and magnetic fields for backgrounds where the boundary is four or three dimensional and satisfies homogeneity and isotropy. We find a frame-independent expression for the DC conductivity tensor. For particular backgrounds we recover previous results on  holographic metals and strange metals.}


\preprint{ICTP-SAIFR/2013-002}

\begin{document}

\section{Introduction}

Holography provides a new approach for investigating strongly coupled field theories. The most important case is given by the AdS/CFT correspondence \cite{Maldacena:1997re} that maps gauge theories with conformal symmetry to String/M theories compactified into Anti-de-Sitter spacetimes. The AdS/CFT correspondence was conjectured by Maldacena after investigating N coincident branes in  String/M theory in the large-N limit. At low energies some of these configurations describe conformal gauge theories at weak coupling while at strong coupling they lead to supergravity solutions involving Anti-de-Sitter spacetimes.

For example, the holographic dual of  ${\cal N}=4$  $U(N)$ Super Yang-Mills theory in 4-d is Type IIB String Theory in $AdS_{5} \times S^{5}$. This duality arises from a configuration of N coincident D3-branes in the large-N limit. This example is very interesting for many theoretical reasons (for instance integrability) and provides new insights into the old problem of non-perturbative QCD.
Another interesting example is the ${\cal N}=6$ k-level $U(N) \times U(N)$ Super Chern-Simons gauge theory in 3-d, whose holographic dual is M-theory in $AdS_4 \times S^7/Z_k$. In this case the duality arises after considering a configuration of M2-branes in M theory. This example has inspired recent holographic models for condensed matter physics.

An interesting problem in condensed matter systems is to describe charge transport in strongly coupled theories. In holography, this is achieved by the inclusion of fermionic charge carriers in the AdS/CFT correspondence. A very simple method of including fermions in the AdS/CFT correspondence consists of the inclusion of $N_f$  Dp-branes in the probe limit $N_f \ll N$. This was originally proposed in the 4-d case \cite{Karch:2002sh} where $N_f$ D7-branes are included on top of $N$ D3-branes. The probe limit $N_f \ll N$ guarantees that the backreaction of the D7-branes on the bulk geometry can be ignored. The fermions are interpreted as quarks in the 4-d field theory whose dual are strings stretching from the D3-branes to the D7-branes.

In this paper we consider the problem of probe  Dp-branes in an arbitrary static background whose metric depends only on the holographic radial coordinate $z$. We provide a simple and elegant method to write the Dirac-Born-Infeld (DBI) equations that describe the dynamics of the Dp-brane. Using these results we calculate the electromagnetic currents through holographic methods for a general gauge field configuration. We find a nice expression for the case when the gauge field strengths on the brane depends only on the radial coordinate $z$. As an application, we consider the case of constant electric and magnetic fields and calculate the DC conductivity tensor when the boundary is four and three dimensional and satisfies homogeneity and isotropy. This is done following the Karch-O'Bannon's procedure \cite{Karch:2007pd,O'Bannon:2007in} (see also \cite{Ammon:2009jt}) which is based on the reality of the brane action. We check our results for particular backgrounds, namely the D3-D7 model \cite{Karch:2002sh} and the Lifshitz background \cite{Kachru:2008yh,Danielsson:2009gi,Mann:2009yx,Bertoldi:2009vn,Balasubramanian:2009rx}. Our results for the D3-D7 model are consistent with those of \cite{Karch:2007pd,O'Bannon:2007in,Ammon:2009jt} and are interpreted as holographic metals while in the Lifshitz case the conductivity is  consistent with  \cite{Hartnoll:2009ns} and can be interpreted in terms of strange metals. Our results are also consistent with a recent method proposed in \cite{Kim:2011qh,Kim:2011zd}, which is based on the properties of an effective metric, called open string metric, that arises from the DBI action.

Our paper ends with some conclusions and an appendix describing the Drude model and the effect of a 5-d Chern-Simons term on the DBI equations for constant electric and magnetic fields.

\section{Dp-brane dynamics and holographic currents}

Consider the dynamics of a probe Dp-brane in an
arbitrary static background of the form
\beqa
ds^2 = a(z) g_{mn}(z) dx^m dx^n + b(z) d \Sigma^2 \,  ,
\eeqa
where $g_{mn}$ describes a ${\cal D} + 1$ spacetime that includes the
radial coordinate $z$. We will be interested in the cases ${\cal D}=4$ and ${\cal D}=3$ although the results in this section are general.

We assume for simplicity that $\Sigma$ is a compact space. We consider the case where the Dp-brane fills the spacetime of the internal coordinates $x^m$ and wrapping a submanifold $\Omega$ of dimension $p - {\cal D}$ inside $\Sigma$. In a general class of Dp-brane models, the submanifold $\Omega$ is specified by fixing some coordinates in $\Sigma$ to constant values while one of the coordinates, say $\theta$,  depends on the radial coordinate $z$.

The induced metric on the Dp-brane takes the form
\beqa
ds_{\ss Dp}^2 = a(z) g_{mn}(z) dx^m dx^n + b(z) d \Omega^2 =: G_{MN} dx^M dx^N  \, ,
\eeqa
where $g_{zz}$ now includes a contribution arising from $d \theta^2 = \theta'^2(z) dz^2$.

\medskip

\subsection{The DBI action}

\medskip

The dynamics of a single Dp brane is described by the abelian  DBI action
\beqa
S_{\rm DBI} = - \mu_p \int d^{{\cal D}} x dz \, d^{p-{\cal D}} \Omega \,  e^{- \phi } \sqrt{- \det (G_{MN} + 2 \pi \alpha' F_{MN} )} \, , \label{DBI1}
\eeqa
where $G_{MN}$ is the induced metric on the probe brane, $F_{MN}$ is the world volume gauge field strength and
\beqa
\mu_p = (2 \pi)^{-p} \alpha'^{- \frac{p+1}{2}}
\eeqa
is the brane tension.

We assume for simplicity, that the gauge fields have nonzero components in the $x^m$ directions only and these components do not depend on the coordinates of $\Omega$. Thus, after integrating over the compact submanifold $\Omega$ we obtain the ${\cal D}+1$ dimensional effective action that can be written as
\beqa
 S_{\rm DBI} = - \int\!{\rm d}^{\cal D} x {\rm d}z \, \gamma
(z) \sqrt{- E}  \label{DBI2}
\eeqa
where $E$ is the determinant of the tensor defined by
\beqa
E_{mn} := g_{mn} + \beta(z) {\cal F}_{mn} \, ,
\eeqa
and
\beqa
\beta (z) &=& \frac{2 \pi \alpha'}{a(z)} \, , \cr
\gamma (z) &=&  \mu_p \, V_{\Omega}  \,   e^{-\phi } (a(z))^{\frac{{\cal D} +1 }{2}} (b(z))^{\frac{p-{\cal D}}{2}}  \,. \label{parameters1}
\eeqa
Here we have used the notation $V_\Omega$ to refer to the volume of the compact submanifold $\Omega$. Note that ${\cal F}_{mn}$ denotes the components of the field strength in the $x^m$ directions.

\medskip

\subsection{DBI equations and holographic currents}

\medskip

The relevant fields in the effective action (\ref{DBI2}) are the ${\cal D}+1$ dimensional gauge field  ${\cal A}_m$  and the scalar field $ \theta (z) $. Therefore the variation of the action (\ref{DBI2}) takes the form
\beqa
\delta S_{\rm DBI} = \delta S^{\rm Bulk}_{\rm DBI} + \delta S^{\rm Bdy}_{\rm DBI}
\eeqa
where
\beqa
\delta S^{\rm Bulk}_{\rm DBI} &=& \int d^{\cal D} x dz  \Big \{ \partial_m \left ( \frac{\gamma \beta}{2} \sqrt{-E} \tilde E^{(\ell, m)} \right) \delta {\cal A}_\ell \cr
 &+& \left [ \partial_z \left ( \frac{\gamma}{2} \sqrt{-E} \tilde E^{zz} \frac{\partial g_{zz}}{\partial \theta'} \right )
-  \sqrt{-E} \frac{\partial \gamma}{\partial \theta} \right ]\delta \theta
\Big \}  \, , \label{deltaSBulk} \\
\delta S^{\rm Bdy}_{\rm DBI}  &=&  - \int d^{\cal D} x dz  \Big \{ \partial_m \left [ \left ( \frac{\gamma \beta}{2} \sqrt{-E} \tilde E^{(\ell,m)} \right)
\delta {\cal A}_\ell \right ] + \partial_z \left [ \frac{\gamma}{2} \sqrt{-E} \tilde E^{zz} \frac{\partial g_{zz}}{\partial \theta'} \delta \theta \right ] \Big \}  \,, \cr
&& \label{deltaSBdy}
\eeqa
and $\tilde E^{(\ell,m)}= \tilde E^{\ell m} - \tilde E^{m \ell}$ being $\tilde E^{\ell m}$ the inverse of $E_{\ell m}$.

Since the field variations $\delta A_\ell$ and $\delta \theta$ are arbitrary, the bulk term (\ref{deltaSBulk}) leads to the DBI equations
\beqa
\partial_m \left ( \frac{\gamma \beta}{2} \sqrt{-E} \tilde E^{(\ell, m)} \right) &=& 0 \, ,  \label{DBIeq1} \\
\partial_z \left ( \frac{\gamma}{2} \sqrt{-E} \tilde E^{zz} \frac{\partial g_{zz}}{\partial \theta'} \right )
-  \sqrt{-E} \frac{\partial \gamma}{\partial \theta} &=& 0 \,. \label{DBIeq2}
\eeqa

A natural prescription for the boundary electromagnetic currents in holography is the following
\beqa
j^\mu = \frac{\delta S}{\delta {\cal A}_\mu} \vert_{z = z_{\rm Bdy}} \,.
\eeqa
In our case the only contribution comes from the first term in the boundary term (\ref{deltaSBdy}), so we find
\beqa
j^\mu = - \lim_{z \to z_{\rm Bdy}} \left [ \frac{ \gamma \beta}{2}
\sqrt{-E} \tilde E^{(\mu , z)} \right ] \,. \label{DBIji}
\eeqa

The relevant equation for the gauge fields are (\ref{DBIeq1}). Decomposing these equations into $\ell = (z,0,i)$ components we obtain \beqa
&& \partial_0 \left ( \frac{\gamma \beta}{2} \sqrt{-E} \tilde E^{(z,0)} \right)
+ \partial_i \left ( \frac{\gamma \beta}{2} \sqrt{-E} \tilde E^{(z,i)} \right) = 0 \, , \label{DBIeqAz} \\
&&  - \partial_z \left ( \frac{\gamma \beta}{2} \sqrt{-E} \tilde E^{(z,0)} \right)
+ \partial_i \left ( \frac{\gamma \beta}{2} \sqrt{-E} \tilde E^{(0,i)} \right) = 0 \, ,  \label{DBIeqA0}\\
&& -\partial_z \left ( \frac{\gamma \beta}{2} \sqrt{-E} \tilde E^{(z,i)} \right)
- \partial_0 \left ( \frac{\gamma \beta}{2} \sqrt{-E} \tilde E^{(0,i)} \right) + \partial_j \left ( \frac{\gamma \beta}{2} \sqrt{-E} \tilde E^{(i,j)} \right) = 0 \,.  \label{DBIeqAi}
\eeqa
In the case where the gauge field strength tensor ${\cal F}_{mn}$ depends only on the radial coordinate $z$, the equation (\ref{DBIeqAz}) is automatically satisfied while the others can be integrated out leading to the equation
\beqa
\frac{\gamma \beta}{2} \sqrt{-E} \tilde E^{(z,\mu)} &=&  j^\mu  \,, \label{DBIeqcurrents}
\eeqa
where $j^\mu$ is the holographic current associated with the DBI action, as obtained in (\ref{DBIji}). This is a non-trivial equation involving the field strengths on the l.h.s and the holographic currents on the r.h.s.

\medskip

{\bf The contribution from a $\Theta$-term}

\medskip

In the ${\cal D}=4$ case, the boundary field theory is four dimensional and may include a $\Theta$-term of the form\footnote{A $\Theta$-term may arise from instanton configurations in QCD-like theories and breaks parity symmetry, see for instance \cite{Kharzeev:2009fn}}.
\beqa
S_{\Theta} = \epsilon^{\mu \nu \rho \sigma} \int dt d^3 \overline{x}
\, \Theta  F_{\mu \nu} F_{\rho \sigma} \,, \label{thetaaction}
\eeqa
where $\Theta = \Theta(t, \vec{x})$ is a 4-d axion field and $F_{\mu \nu} = \partial_\mu A_\nu - \partial_\nu A_\mu$ is the four dimensional field strength.

Integrating by parts (\ref{thetaaction}) we obtain
\beqa
S_{\Theta} = 2 \epsilon^{\mu \nu \rho \sigma} \int dt d^3 \overline{x}  \Big \{ \partial_\mu \left [  \Theta A_\nu F_{\rho \sigma} \right ]
+ A_\mu (\partial_\nu \Theta) F_{\rho \sigma} \Big \}  \,,
\eeqa
This term provides a contribution to the holographic current :
\beqa
\Delta j^\mu = \frac{ \delta S_{\Theta} }{\delta A_\mu}
= 2  \epsilon^{\mu \nu \rho \sigma} (\partial_\nu \Theta) F_{\rho \sigma} \,.
\eeqa

\medskip

\subsection{The method of $\beta$-expansion}

\medskip

The method proposed in this paper consists on a series expansion in the parameter $\beta(z)$ for the relevant quantities appearing on the DBI equations, namely the determinant $E$ and the inverse tensor $\tilde E^{m,n}$.

Consider the matrix representation of the metric $\hat g$ and the field strength $\hat F$. Then the tensor $E_{mn}$ has a matrix representation given by
\beqa
\hat E = \hat g + \beta \hat F = \hat g( 1 + \hat X) \, ,
\eeqa
where we have defined $\hat X := \beta \hat g^{-1} \hat F$. Note that this matrix is matrix linear in $\beta$. Recalling some properties of the determinant we can write $E$ in terms of $\hat X$ :
\beqa
E = \det (\hat E) &=& \det (\hat g)  \det ( 1 + \hat X) \cr
&=& \det (\hat g) \exp \{ \tr [ \ln ( 1 + \hat X) ]\} \,. \label{detexp}
\eeqa
On the other hand, the tensor $E^{m,n}$ corresponds to the inverse matrix $\hat E^{-1}$ and can also be represented in terms of $\hat X$  :
\beqa
\hat E^{-1} = ( 1 + \hat X)^{-1} g^{-1} \, . \label{inverseE}
\eeqa
Now we consider a series expansion in $\hat X$ for the determinant $E$ and the inverse tensor $\tilde E^{m,n}$.
For the determinant we first expand the matrix $ \ln ( 1 + \hat X) $ and take the trace :
\beqa
\tr [ \ln ( 1 + \hat X) ] = - \frac12 \tr (\hat X^2)  - \frac14 \tr (\hat X^4 ) - \frac16 \tr (\hat X^6) + \dots \, , \label{Yexp}
\eeqa
where we used the identity $\tr (\hat X^n ) = 0$ for odd values of $n$. Substituting (\ref{Yexp}) into (\ref{detexp}) we get
\beqa
E &=& \det (\hat g) \Big \{ 1 - \frac12 \tr (\hat X^2 ) - \frac14 \tr (\hat X^4) + \frac18 [ \tr (\hat X^2) ]^2 \cr
&-& \frac16 \tr ( \hat X^6)  + \frac18 [ \tr ( \hat X^2 ) ] [ \tr ( \hat X^4 ) ]
- \frac{1}{48} [ \tr ( \hat X^2 ) ]^3  + \dots \Big \} \, .
\eeqa
 The traces appearing in the expansion can be written in a tensorial form
\beqa
\tr (\hat X^2 ) &=& - \beta^2 {\cal F}^{mn} {\cal F}_{mn} \, , \cr
\tr (\hat X^4) &=&  \beta^4 {\cal F}^{mn} {\cal F}_{np} {\cal F}^{pq} {\cal F}_{qm} \, , \cr
\tr (\hat X^6) &=&  \beta^6 {\cal F}^{mn} {\cal F}_{np} {\cal F}^{pq} {\cal F}_{qr} {\cal F}^{rs} {\cal F}_{sm} \, , \quad \dots \label{betaf}
\eeqa
Then we conclude that the determinant has the following $ \beta$-expansion
\beqa
E &=& \det (\hat g) \Big \{ 1 + \frac{\beta^2}{2} {\cal F}^{mn} {\cal F}_{mn} + \frac{\beta^4}{4 !}  {\cal F}^{mnpq}  {\cal F}_{mnpq} \cr
&+& \frac{\beta^6}{6!} {\cal F}^{mnpqrs} {\cal F}_{mnpqrs} + \dots \Big \} \,, \label{det1}
\eeqa
where
\beqa
 {\cal F}_{mnpq} &:=& {\cal F}_{mn} {\cal F}_{pq} - {\cal F}_{mp} {\cal F}_{nq} + {\cal F}_{mq} {\cal F}_{np}  \, , \cr
{\cal F}_{mnpqrs} &:=& {\cal F}_{mn}  {\cal F}_{pqrs} - {\cal F}_{mp}  {\cal F}_{nqrs} + {\cal F}_{mq}  {\cal F}_{nprs} \cr
&-& {\cal F}_{mr}  {\cal F}_{npqs} + {\cal F}_{ms} {\cal F}_{npqr} \, , \dots \label{fmn}
\eeqa
It is easy to check that the covariant tensors ${\cal F}_{mnpq}$ and ${\cal F}_{mnpqrs}$ are antisymmetric in all the index\footnote{The antisymmetry of these tensors implies that they admit a representation in terms of Levi-Civita symbols. We will not use that representation in this paper, though.}.

 The expansion (\ref{det1}) involves only even powers of $\beta$. It can also be truncated according to the dimension ${\cal D}+1$ of the spacetime relevant to the effective action. This is due to the antisymmetry of the tensors appearing in the expansion. Consider for instance  the cases ${\cal D}=4$ or ${\cal D} =3$ where the effective bulk theory is five or four dimensional respectively. In those cases the antisymmetry of the tensor ${\cal F}_{mnpqrs}$ defined in (\ref{fmn}) implies that this tensor vanishes. This is also true for the tensors appearing at higher order in $\beta$.  Therefore for ${\cal D}=4$ and ${\cal D}=3$ the $\beta$-expansion (\ref{det1}) reduces to a polynomial or order $\beta^4$.

Now consider the matrix representation (\ref{inverseE}) of the inverse tensor $E^{m,n}$. Expanding  $  ( 1 + \hat X)^{-1}$  in a geometric series and  using (\ref{betaf}) we get the following expansion :
\beqa
\tilde E^{mn} &=&  g^{mn} - \beta {\cal F}^{mn} + \beta^2 {\cal F}^m_{\, \, \,  p} {\cal F}^{p n} - \beta^3 {\cal F}^{mp} {\cal F}_{pq} {\cal F}^{qn} \cr
&+& \beta^4 {\cal F}^m_{\, \, \, p} {\cal F}^{pq} {\cal F}_{qr} {\cal F}^{rn} - \beta^5 {\cal F}^{mp} {\cal F}_{pq} {\cal F}^{qr} {\cal F}_{rs} {\cal F}^{sn} + \dots \, . \label{exptildeEmn}
\eeqa
We are also interested in the antisymmetric tensor $\tilde E^{(m,n)} = \tilde E^{mn} - \tilde E^{nm}$. The terms in the expansion (\ref{exptildeEmn}) that contain even powers of $\beta$ are symmetric in $m \leftrightarrow n$. Therefore, the tensor $\tilde E^{(m,n)}$ takes the form
\beqa
\tilde E^{(m,n)} =  - 2 \beta {\cal F}^{mn}  - 2 \beta^3 {\cal F}^{mp} {\cal F}_{pq} {\cal F}^{qn}
- 2 \beta^5 {\cal F}^{mp} {\cal F}_{pq} {\cal F}^{qr} {\cal F}_{rs} {\cal F}^{sn} + \dots \, . \label{expantisymEmn}
\eeqa

If we multiply the expansion (\ref{expantisymEmn}) with the  determinant given in (\ref{det1}) we obtain the expansion
\beqa
E \tilde E^{(m,n)} = \det (\hat g) \Big \{  - 2 \beta {\cal F}^{mn} - \beta^3 {\cal F}_{pq} {\cal F}^{mnpq}
- \frac{\beta^5}{4} {\cal F}_{pq} {\cal F}_{rs}  {\cal F}^{mnpqrs} + \dots \Big \} \,. \label{detEE}
\eeqa

\medskip

{\bf The DBI effective action and DBI equations}

\medskip

The $\beta$-expansion of the determinant $E$ allows us to find a useful expansion for the effective DBI action given in (\ref{DBI2}). In (\ref{DBI2}) the lagrangian density is proportional to $\sqrt{-E}$. Using the $\beta$-expansion in (\ref{det1}) we find that
\begin{equation}
\sqrt{- E} = \sqrt{-g} Q(z,x) \, , \label{det2}
\end{equation}
with $Q(z,x)$ given by
\beqa
Q(z,x) = \Big \{ 1 + \frac{\beta^2}{2} {\cal F}^{mn} {\cal F}_{mn} + \frac{\beta^4}{4 !}  {\cal F}^{mnpq} {\cal F}_{mnpq} + \frac{\beta^6}{6!} {\cal F}^{mnpqrs} {\cal F}_{mnpqrs} + \dots \Big \}^{1/2} \,.
\eeqa

As a consequence of (\ref{det2}) and (\ref{detEE}) we find that
\beqa
\sqrt{-E} \tilde E^{(m,n)} = - 2 \beta \frac{ \sqrt{-g} }{Q(z,x)} \Big \{   {\cal F}^{mn} + \frac{\beta^2}{2} {\cal F}_{pq} {\cal F}^{mnpq}
+ \frac{\beta^4}{8} {\cal F}_{pq} {\cal F}_{rs}  {\cal F}^{mnpqrs} + \dots \Big \} \,.
\label{usefulexp}
\eeqa
This is a very useful expansion. In the general case, where the field strengths depend not only on $z$ but also on time and spatial coordinates, we can use (\ref{usefulexp}) in (\ref{DBIeq1}) to rewrite the DBI equations as
\beqa
\partial_m \Big \{ \gamma \beta^2 \frac{ \sqrt{-g} }{Q(z,x)}
\left [ {\cal F}^{\ell m} + \frac{\beta^2}{2} {\cal F}_{pq}  {\cal F}^{\ell m p q}
 + \frac{\beta^4}{8} {\cal F}_{pq} {\cal F}_{rs}  {\cal F}^{\ell m p q r s}
 + \dots \right ] \Big \} = 0 \,. \label{DBIexpanded}
\eeqa
The DBI equations written in the form (\ref{DBIexpanded}) show explicitly the expansion in $\beta$ which is equivalent to an expansion in the string length squared $\alpha'$. This is useful if we want to truncate the DBI equations in the limit of small $\beta$. In many brane models  the product $\gamma \beta^2$ is independent of $z$ so that in the limit of small $\beta$ the DBI equations reduce to Maxwell equations in a curved spacetime with metric $g_{mn}(z)$.

The traditional way of solving DBI equations is to choose first a gauge field ansatz and then work with an effective determinant. The advantage of (\ref{DBIexpanded}) is that they are valid for a general class of  gauge field configurations, including time and space dependent field strengths.

As a simple application of the method described above, we will consider the case the field strengths $ {\cal F}_{mn}$ depend only on the radial coordinate $z$. This case is relevant to describe the physics of constant electric and magnetic fields in the boundary. In this case we can integrate out (\ref{DBIexpanded}) and obtain
\beqa
j^\mu = - \gamma \beta^2 \frac{ \sqrt{-g} }{Q(z)} \Big \{   {\cal F}^{z \mu} + \frac{\beta^2}{2} {\cal F}_{\nu \rho}  {\cal F}^{z \mu \nu \rho }
+ \frac{\beta^4}{8} {\cal F}_{\nu \rho} {\cal F}_{\lambda \sigma}  {\cal F}^{z \mu \nu \rho \lambda \sigma} + \dots \Big \} \label{currentDBIeq}
\eeqa
This equation is equivalent to (\ref{DBIeqcurrents}) and will be used in the next section to calculate the conductivity in the presence of constant electric and magnetic fields.

\section{DC conductivity in $3+1$ dimensions}

Now we consider the problem of charge transport in a $3+1$ dimensional strongly coupled field theory in the presence of constant electric and magnetic fields. The way of describing this system in holography is by considering a five dimensional effective DBI action of the form (\ref{DBI1}) with a gauge field of the form :
\beqa
{\cal A}_0(z) &=& f_0(z) \quad , \quad \vec {\cal A}(z,x) = \vec{f}(z) -\vec{E} t + \frac12 \vec{B} \times \vec{x}   \,. \label{DCansatz}
\eeqa
For simplicity, we also assume that the metric is homogeneous and isotropy in the spatial directions $\vec{x}$ in the sense that $g_{11} = g_{22} = g_{33} := g_{xx}(z)$.

As discussed in the previous section, the DBI equations in the case where the field strengths depend on the radial coordinate $z$ only can be written as (\ref{currentDBIeq}). For the ansatz (\ref{DCansatz}) the equations (\ref{currentDBIeq}) take the form
\beqa
- \frac{Q(z)}{\gamma \beta^2} j^0 &=&  \sqrt{-g}  g^{zz} g^{00} \left \{ \left [ 1 + \beta^2 (g^{xx})^2 |\vec{B}|^2 \right ]  \partial_z f_0
 + \beta^2 (g^{xx})^2 (\vec{E} \times \vec{B}) \cdot (\partial_z \vec{f}) \right \}  \, , \cr
&& \label{j0eq} \\
- \frac{\gamma \beta^2}{Q(z)} \vec{j} &=&  \sqrt{-g} g^{zz} g^{xx}  \Big \{  \left [ 1 + \beta^2 g^{00} g^{xx} |\vec{E}|^2 \right ]  \partial_z \vec{f} \cr
&-& \beta^2 g^{00} g^{xx} \left [  ( \vec{E} \cdot \partial_z \vec{f}) \vec{E} - (\partial_z f_0) \vec{E} \times \vec{B} \right ]
+ \beta^2 (g^{xx})^2 (\vec{B} \cdot \partial_z \vec{f}) \vec{B} \Big \}  \label{jieq}
\eeqa
where
\beqa
Q(z)  &\equiv&  \Big \{ Q_0(z) + \beta^2  g^{zz} g^{00} \left [ 1 + \beta^2 (g^{xx})^2 |\vec{B}|^2 \right ] (\partial_z f_0)^2  \nonumber \\
&+& \beta^2 g^{zz} g^{xx} \left [ 1 + \beta^2 g^{00} g^{xx} |\vec{E}|^2 \right ] |\partial_z \vec{f}|^2    \cr
&-& \beta^4 g^{zz} g^{00} (g^{xx})^2 \left [-   2 (\partial_z f_0) (\vec{E} \times \vec{B}) \cdot (\partial_z \vec{f})   + (\vec{E} \cdot \partial_z \vec{f})^2 \right ]  \nonumber \\
&+& \beta^4 g^{zz} (g^{xx})^3 \left ( \vec{B} \cdot \partial_z \vec{f} \right )^2 \Big \}^{1/2} \, , \label{defQ}
\eeqa
and
\beqa
Q_0(z) = 1 + \beta^2 \left [ g^{00} g^{xx} |\vec{E}|^2 + (g^{xx})^2 |\vec{B}|^2 \right ]
+ \beta^4 g^{00} (g^{xx})^3  (\vec{E} \cdot \vec{B})^2 \,. \label{defQ0}
\eeqa
Note that $Q_0(z)$ does not depend on the functions $f_0(z)$ and $\vec{f}(z)$.

Taking the scalar product of (\ref{jieq}) with $\vec{E}$, $\vec{B}$ and $\vec{E} \times \vec{B}$ respectively we get the following equations
\beqa
\frac{Q(z)}{\gamma \beta^2} \vec{E} \cdot \vec{j}  &=& - \sqrt{-g} g^{zz} g^{xx} \Big \{  (\vec{E} \cdot \partial_z \vec{f})
+ \beta^2  (g^{xx})^2 (\vec{E} \cdot \vec{B}) (\vec{B} \cdot \partial_z \vec{f}) \Big \} \, , \label{Ejeq} \\
\frac{Q(z)}{\gamma \beta^2} \vec{B} \cdot \vec{j}  &=& - \sqrt{-g} g^{zz} g^{xx} \Big \{ \left [ 1 + \beta^2 g^{00} g^{xx} |\vec{E}|^2 + \beta^2 (g^{xx})^2 |\vec{B}|^2 \right ]  (\vec{B} \cdot \partial_z \vec{f}) \cr
&-& \beta^2 g^{00} g^{xx} (\vec{E} \cdot \vec{B}) (\vec{E} \cdot \partial_z \vec{f}) \Big \} \, , \label{Bjeq} \\
\frac{Q(z)}{\gamma \beta^2} (\vec{E} \times \vec{B}) \cdot \vec{j}  &=& \sqrt{-g} g^{zz} g^{xx} \Big \{ - \left [ 1 + \beta^2 g^{00} g^{xx} |\vec{E}|^2 \right ]( \vec{E} \times \vec{B}) \cdot (\partial_z \vec{f})  \cr
&-& \beta^2 g^{00} g^{xx} |\vec{E} \times \vec{B}|^2 \partial_z f_0 \Big \} \, . \label{EBjeq}
\eeqa

\medskip

\subsection{Solutions of the DBI equations}

\medskip

We can solve the system (\ref{Ejeq}) and (\ref{Bjeq}) as a function of $Q(z)$. The result is
\beqa
\sqrt{-g} g^{zz} g^{xx} \vec{B} \cdot \partial_z \vec{f}  &=& - \frac{Q(z)}{\gamma \beta^2 Q_0(z)}  \Big \{
\vec{B} + \beta^2 g^{00} g^{xx} (\vec{E} \cdot \vec{B}) \vec{E} \Big \} \cdot \vec{j} \cr
&=:& \frac{Q(z)}{\gamma \beta^4 Q_0(z)} j_B \, , \\
\sqrt{-g} g^{zz} g^{xx} \vec{E} \cdot \partial_z \vec{f}  &=& - \frac{Q(z)}{\gamma \beta^2 Q_0(z)} \left [ Q_0(z) \vec{j} \cdot \vec{E} - (g^{xx})^2 (\vec{E} \cdot \vec{B}) j_B \right ] \cr
&=: & \frac{Q(z)}{\gamma \beta^4 Q_0(z)} j_E \,.
 \eeqa

Similarly, we can solve (\ref{j0eq}) and (\ref{EBjeq}) as a function of $Q(z)$. The solutions are given by
\beqa
\sqrt{-g} g^{zz} g^{xx} (\vec{E} \times \vec{B}) \cdot \partial_z \vec{f}  &=& - \frac{Q(z)}{\gamma \beta^2 Q_0(z)} \Big \{
 - \beta^2 (g^{xx})^2 |\vec{E} \times \vec{B}|^2 j^0 \cr
 &+&  \left [ 1 + \beta^2 (g^{xx})^2 |\vec{B}|^2 \right ] (\vec{E} \times \vec{B}) \cdot \vec{j}
\Big \}  \cr
&=:& \frac{Q(z)}{\gamma \beta^5 Q_0(z)}  j_{EB} \, ,
\eeqa
\beqa
\sqrt{-g} g^{zz} g^{00}   \partial_z f_0  &=& \frac{Q(z)}{\gamma \beta^2 Q_0(z)}  \Big \{  - \left [ 1 + \beta^2 g^{00} g^{xx} |\vec{E}|^2 \right ] j^0
+  \beta^2 g^{00} g^{xx} (\vec{E} \times \vec{B}) \cdot \vec{j}  \Big \}  \cr
&=:& \frac{Q(z)}{\gamma \beta^3 Q_0(z)} \tilde j_0 \, .  \label{f0almfinal}
\eeqa

Using these results in (\ref{jieq}) we get
\beqa
\sqrt{-g} g^{zz} g^{xx} \left [ 1 + \beta^2 g^{00} g^{xx} |\vec{E}|^2 \right ] \partial_z \vec{f} &=& \frac{Q(z)}{\gamma \beta^2 Q_0(z)} \Big \{ - Q_0(z) \vec{j}
+  g^{00} g^{xx}  j_E \vec{E} \cr
&-&  \beta (g^{xx})^2  \tilde j_0 \vec{E} \times \vec{B}
- (g^{xx})^2 j_B \vec{B} \Big \} \, . \label{fialmfinal}
\eeqa
Finally, using the solutions in the expansion for $Q(z)$ ( eq. \ref{defQ}) we find a solution for $Q(z)$ :
{\small \beqa
Q(z)
&=& \frac{\gamma \beta^2 Q_0(z)}{\sqrt{ Q_0(z) \chi (z) +  \frac{(g^{xx})^3}{\left [ 1 + \beta^2 g^{00} g^{xx} |\vec{E}|^2 \right ]}  \tilde j_0^2
- \frac{ g^{00} (g^{xx})^4 }{\left [ 1 + \beta^2 g^{00} g^{xx} |\vec{E}|^2 \right ]} j_B^2}} \, , \label{Qonshell}
\eeqa }
where
\beqa
\chi(z) :=  \gamma^2 \beta^4
+ \frac{\beta^2 g^{00} (g^{xx})^2 }{\left [ 1 + \beta^2 g^{00} g^{xx} |\vec{E}|^2 \right ]}
\left [ |\vec{j}|^2 + \beta^2 g^{00} g^{xx} (\vec{j} \cdot \vec{E})^2 \right ]  \,. \label{defchi}
\eeqa
Note that $Q(z)$ has now been expressed in terms of the physical quantities of the system, namely the charge density $j^0$, the electromagnetic current $\vec{j}$ and the electric and magnetic fields $\vec{E}$ and $\vec{B}$.

\medskip

\subsection{The Karch-O'Bannon's procedure and the conductivity tensor in a general frame}

\medskip
The onshell DBI action can be written as
\beqa
S_{\rm DBI}^{\rm o.s} = - \int d^4 x dz \gamma(z) \sqrt{-g} Q(z) \, ,
\eeqa
with $Q(z)$ given by (\ref{Qonshell}).

The Karch-O'Bannon's procedure proposed in \cite{Karch:2007pd} and further developed in \cite{O'Bannon:2007in,Ammon:2009jt} consists in imposing the reality constraint on the onshell DBI action.
In a general class of backgrounds the metric blows up at the boundary while the component $g_{tt}$ vanishes at the horizon. Under these circumstances, it is not difficult to see from (\ref{defQ0}) and (\ref{defchi}) that the functions $Q_0(z)$ and $\chi(z)$ should  have a zero at a point $z=z_\ast$ and $z = z_\chi$ respectively. Using the identity
\beqa
\left [ 1 + \beta^2 g^{00} g^{xx} |\vec{E}|^2 \right ]
\left [ 1 + \beta^2 (g^{xx})^2 \frac{(\vec{E} \cdot \vec{B})^2 }{\vec{E}^2} \right ] = Q_0(z) - \beta^2 (g^{xx})^2 \frac{ |\vec{E} \times \vec{B}|^2}{|\vec{E}|^2} \, ,
\eeqa
we conclude that $1 + \beta^2 g^{00} g^{xx} |\vec{E}|^2 < 0$ at $z = z_\ast$. Therefore, in order to avoid an imaginary action we find from (\ref{Qonshell}) that $\tilde j_0$ and $j_B$ must vanish at $z = z_\ast$. Since $1 + \beta^2 g^{00} g^{xx} \vec{E}^2 $ is also positive at the boundary it should have a zero at $z = z_{E^2}$, which is closer to the boundary than $z_\ast$. Analyzing all the possible localizations of $z_\chi$  with respect to $z_{\ast}$ and $z_{E^2}$ leads to the conclusion that $z_\chi = z_\ast$ (for more details see \cite{Ammon:2009jt}).

In summary, we find from the reality constraint on the action that at some point  $z = z_\ast$ the following equations must be satisfied :
\beqa
Q_0 &=& 1 + \beta^2 g^{00} g^{xx} |\vec{E}|^2 + \beta^2(g^{xx})^2 |\vec{B}|^2 + \beta^4 g^{00} (g^{xx})^3  (\vec{E} \cdot \vec{B})^2 = 0 \, , \label{constraint0} \\
j_B &=& - \beta^2 \left [ \vec{j} \cdot \vec{B} + \beta^2 g^{00} g^{xx} (\vec{E} \cdot \vec{B}) \vec{j} \cdot \vec{E} \right ] = 0 \, , \label{constraint1} \\
\tilde j_0 &=& \beta \left \{
- \left [ 1 + \beta^2 g^{00} g^{xx} |\vec{E}|^2 \right ] j^0
+ \beta^2 g^{00} g^{xx} \vec{j} \cdot ( \vec{E} \times \vec{B} ) \right \} = 0 \, , \label{constraint2} \\
\chi &=& \gamma^2 \beta^4
+ \frac{\beta^2 g^{00} (g^{xx})^2 }{\left [ 1 + \beta^2 g^{00} g^{xx} |\vec{E}|^2 \right ]}
\left [ |\vec{j}|^2 + \beta^2 g^{00} g^{xx} (\vec{j} \cdot \vec{E})^2 \right ] = 0 \, . \label{constraint3}
\eeqa
From eqs. (\ref{constraint0}), (\ref{constraint1}) and (\ref{constraint2}) we get
\beqa
1 +  \beta^2 (g^{xx})^2 |\vec{B}|^2 &=&
- \beta^2 g^{00} g^{xx} \left [ |\vec{E}|^2 + \beta^2 (g^{xx})^2 (\vec{E} \cdot \vec{B})^2 \right ] \,  ,  \label{Q0der} \\
\vec{j} \cdot (\vec{E} \times \vec{B}) &=&
\frac{\beta^2 (g^{xx})^2 |\vec{E} \times \vec{B}|^2 j^0}
{\left [ 1 + \beta^2 (g^{xx})^2 |\vec{B}|^2 \right ]} \, , \label{jEBj0} \\
\vec{j} \cdot \vec{B} &=&  - \beta^2 g^{00} g^{xx} (\vec{E} \cdot \vec{B}) \vec{j} \cdot \vec{E} \, \label{jEjB}.
\eeqa
The spatial current $\vec{j}$ can be decomposed into projections on $\vec{E}$, $\vec{B}$ and $\vec{E} \times \vec{B}$ as
\beqa
\vec{j} &=& \frac{1}{|\vec{E} \times \vec{B}|^2} \Big \{ \left [ (\vec{j} \cdot \vec{E}) |\vec{B}|^2 - (\vec{j} \cdot \vec{B}) (\vec{E} \cdot \vec{B}) \right ] \vec{E} \cr
&+& \left [ (\vec{j} \cdot \vec{B}) |\vec{E}|^2 - (\vec{j} \cdot \vec{E}) (\vec{E} \cdot \vec{B})  \right ] \vec{B}
+ \vec{j} \cdot (\vec{E} \times \vec{B}) \vec{E} \times \vec{B} \Big \} \, . \label{jdecomp}
\eeqa

Using (\ref{jEBj0}),  (\ref{jEjB}) and (\ref{jdecomp}) we get
\beqa
\vec{j} &=& - \frac{\beta^2 g^{xx}}
{ \left [ 1 + \beta^2 (g^{xx})^2 |\vec{B}|^2 \right ] }
\Big \{ g^{00} (\vec{j} \cdot \vec{E})
\left [ \vec{E} + \beta^2 (g^{xx})^2 (\vec{E} \cdot \vec{B}) \vec{B} \right ] \cr
&-& g^{xx} j^0 \vec{E} \times \vec{B} \Big \} \, . \label{jjE}
\eeqa

Using (\ref{constraint3}) and (\ref{jjE}) we can solve $\vec{j} \cdot \vec{E}$. Assuming $\vec{j} \cdot \vec{E} \ge 0$ we get
\beqa
\vec{j} \cdot \vec{E}  &=&  - \frac{g_{00}}{\beta^2} \sqrt{ (g_{xx})^3 \left [ 1 + \beta^2 (g^{xx})^2 |\vec{B}|^2 \right ] \gamma^2 \beta^4 + \beta^2 (j^0)^2 }\,. \cr
&& \label{soljE}
\eeqa
Writing the current as $j_i = \sigma_{ij} E_j$ we get the conductivity tensor
\beqa
\sigma_{ij} &=& \frac{  g^{xx} }{ \left [ 1 + \beta^2 (g^{xx})^2 |\vec{B}|^2 \right ] } \Big \{
\left [ \sqrt{ (g_{xx})^3 \left [ 1 + \beta^2 (g^{xx})^2 |\vec{B}|^2 \right ] \gamma^2 \beta^4 + \beta^2 (j^0)^2} \right ] \cr
&\times& \left [ \delta_{ij} + \beta^2 (g^{xx})^2 B_i B_j \right ]
+ \beta^2 g^{xx} j^0 \epsilon_{ijk} B_k \Big \} \,. \label{condtensor3d}
\eeqa
This result is consistent with the one obtained in \cite{Karch:2007pd,Ammon:2009jt} for the D3-D7 brane model. In that case the induced metric can be written as
\beqa
\frac{ds_{\ss D7}^2}{R^2} &=& - \frac{1}{z^2} \frac{\left ( 1 - \frac{z^4}{z^4_H} \right )^2}{\left ( 1 + \frac{z^4}{z^4_H} \right )}dt^2   + \frac{1}{z^2} \left ( 1 + \frac{z^4}{z^4_H} \right ) d \vec{x}^2
+ \left [ \frac{1}{z^2} + (\partial_z \theta)^2 \right ] dz^2  \cr
&& \cr
&+& \cos^2 \theta(z) d \Omega_3^2 \, .
\eeqa
The black brane horizon is related to the field theory temperature by $z_H = \sqrt{2}/(\pi T)$. The dilaton is just  $e^{-\phi} = g_s^{-1}$ and the AdS radius is given by $R = (4 \pi \lambda )^{1/4} \sqrt{\alpha'}$ with $\lambda = g_s N_c$. Then, comparing to (\ref{DBI2}) we obtain
\beqa
\beta(z) &=& \frac{2 \pi \alpha'}{R^2}  = \sqrt{\pi} \lambda^{-1/2} \, , \cr
\gamma(z) &=& \frac{\mu_7}{g_s} V_{\ss S^3} R^8 \cos^3 \theta(z) = \frac{\lambda N_c}{4 \pi^3} \cos^3 \theta(z)  \, , \cr
g_{xx}(z) &=& \frac{1}{z^2} \left ( 1 + \frac{z^4}{z^4_H} \right ) \,.
\label{D3D7relations}
\eeqa

Our result (\ref{condtensor3d}) is also consistent with that obtained recently in \cite{Kim:2011qh,Kim:2011zd}, using the method of open string metric.

\medskip

\subsection{The Drude limit and holographic metals}

\medskip

As observed in \cite{Karch:2007pd,Ammon:2009jt}, the Drude limit is obtained by neglecting the effects of pair production (large mass limit) and higher powers in the electric field. In our framework, this corresponds to the limit $\gamma \to 0$ and $g_{00} \to 0$. Then (\ref{condtensor3d}) reduces to
\beqa
\sigma^D_{ij} = \frac{j^0}{\mu M \left [ 1 + \frac{|\vec{B}|^2}{\mu^2 M^2} \right ] }
\Big [ \delta_{ij} + \frac{B_i B_j}{\mu^2 M^2} + \epsilon_{ijk} \frac{B_k}{\mu M} \Big ] \,, \label{Drude3d}
\eeqa
where
\beqa
\mu M = \frac{1}{\beta} g_{xx} (z_\ast) \,.
\eeqa
As shown in appendix A, this is exactly the Drude conductivity that describes a metallic behavior for particles with density $j^0$ and mass $M$.
Note that in the case of a black brane the induced metric contains a horizon defined by $g_{00}=0$. Then in the Drude limit the point $z = z_\ast$ coincides with the horizon $z_H$. In particular, for the D3-D7 brane model we get from (\ref{D3D7relations}) :
\beqa
\mu M = 2 \frac{ \sqrt{ \lambda} }{ \sqrt{\pi} z_H^2}
= \frac{\pi}{2} \sqrt{ \bar \lambda} T^2 \, , \label{muMtemp}
\eeqa
where $\bar \lambda = 4 \pi \lambda$.  As discussed in \cite{Karch:2007pd,Ammon:2009jt}, this result can also be obtained by computing the drag force on the charge carriers in the holographic metals.

The temperature dependence in (\ref{muMtemp}) is typical of metals. In the next section we will consider backgrounds with Lifshitz symmetry where the conductivity has a different dependence on temperature.

\section{DC conductivity in 2+1 dimensions}

In $2+1$ dimensions the magnetic field is a scalar so the gauge field ansatz takes the form
\begin{eqnarray}
{\cal A}_0(z) = f_0(z) , &\, \, \, \, \, \, \, \, \, \, \vec{{\cal A}}(z,x) = \vec{f}(z) - \vec{E} t + {1 \over 2} B \epsilon_{ij} x_i \hat x_j \,, \label{ansatz4d}
\end{eqnarray}
where $\hat x_i$ is the unit vector in the spatial directions $i=(1,2)$ and $\epsilon_{ij}$ is the Levi-Civita symbol.
Since the field strengths depend only on the $z$ coordinate we can use (\ref{currentDBIeq}) to get the DBI equations :
\begin{eqnarray}
- j^0 &=& {\gamma \beta^2 \sqrt{-g} \over Q(z)} g^{zz} g^{00} \Big\{\Big[1+ \beta^2 (g^{xx})^2 B^2\Big] \partial_z f_0 + \beta^2 (g^{xx})^2 (\vec{ \tilde E} \cdot \partial_z \vec f ) B\Big\}
\label{DBI3dj0} \\
- \vec j & = & {\gamma \beta^2 \sqrt{-g} \over Q(z)} g^{zz} g^{xx} \Big\{\Big[1 + \beta^2 g^{00} g^{xx} | \vec{E}|^2\Big]\partial_z \vec{f} - \beta^2 g^{00} g^{xx} \Big[(\vec{E} \cdot \partial_z \vec f ) \vec E \nonumber\\& & - \, B (\partial_z f_0) \, \vec{\tilde E} \,  \Big] \Big\} \label{DBI3dji} \, ,
\end{eqnarray}
where $\vec{\tilde E} = \varepsilon_{ij} E_j \hat x_i $ and $Q(z)$ is given by
\beqa
Q(z) &= &\{Q_0(z) + \beta^2 g^{zz} g^{00} [1+ \beta^2 (g^{xx})^2 B^2] (\partial_z f_0)^2 \nonumber \\
&+& \beta^2 g^{zz} g^{xx} [1+\beta^2 g^{00}g^{xx}  |\vec{E}|^2] |\partial_z \vec{f}|^2 \nonumber \\ &- &  \beta^4 g^{zz}g^{00}(g^{xx})^2 [- 2 \partial_z f_0 (\vec{\tilde E} \cdot \partial_z \vec{f}) B + (\vec{E} \cdot \partial_z \vec f )^2]\}^{1/2} \label{1} \, ,
\eeqa
with
\begin{equation}
Q_0(z) = 1 + \beta^2 g^{00} g^{xx} |\vec{E}|^2 + \beta^2 (g^{xx})^2 B^2  \label{Q03d} \,.
\end{equation}

Following a procedure similar to that described in the previous section we solve (\ref{DBI3dj0}) and (\ref{DBI3dji}) in terms of $Q(z)$. We find in this case
\begin{eqnarray}
\sqrt{-g}  g^{zz} g^{00}  \partial_z f_0
&=& {Q(z) \over \gamma \beta^2 Q_0(z)} \Big \{
- \left [ 1 + \beta^2 g^{00} g^{xx} |\vec{E}|^2 \right ] j^0
+ \beta^2 g^{00} g^{xx} B \, \vec{j} \cdot \vec{ \tilde E} \Big \} \cr
& =:&  {Q(z) \over \gamma \beta^3 Q_0(z)} \tilde j_0 \label{13} \, ,
\eeqa
\beqa
\sqrt{-g} g^{zz} g^{xx} \vec E \cdot \partial_z \vec f  &=& - {Q(z) \over \gamma \beta^2 } \vec{j} \cdot \vec{E}  \label{11} \, , \\
\sqrt{-g} g^{zz} g^{xx}  \vec{\tilde E} \cdot \partial_z \vec f
&=& - {Q(z) \over \gamma \beta^2 Q_0(z)} \Big \{
\left [ 1 + \beta^2 g^{zz} g^{xx} B^2 \right ] \vec{j} \cdot \vec{ \tilde E} \cr
&-& \beta^2 (g^{xx})^2 B |\vec{E}|^2 j^0 \Big \} \, ,  \label{12}
\eeqa
\beqa
 \sqrt{-g}g^{zz} g^{xx} \Big[1 + \beta^2 g^{00} g^{xx} |\vec{E}|^2\Big]\partial_z \vec{f} &=& - {Q(z) \over \gamma \beta^2 Q_0(z) } \Big[ \,Q_0(z)\, \vec j \cr
 &+& \beta^2 g^{00} g^{xx} Q_0(z) (\vec{j} \cdot \vec{E}) \vec E \cr
&+&  \beta  (g^{xx})^2\,\tilde j_0 B  \vec{\tilde E}  \Big] \, , \label{17}
\eeqa
Finally, using the results ((\ref{13}), (\ref{11}), (\ref{12}) and (\ref{17}) in (\ref{1}) we find  a solution for $Q(z)$ in terms of the current and electromagnetic fields :
\beqa
Q(z) = \frac{\gamma \beta^2 Q_0(z)}{ \sqrt{ Q_0(z) \chi(z) + \frac{ (g^{xx})^2 }{ \left [ 1 + \beta^2 g^{00}g^{xx} |\vec{E}|^2 \right ] } \tilde j_0^2 }} \,,
\eeqa
where
\beqa
\chi(z) :=
\gamma^2 \beta^4 +{\beta^2 g^{00} g^{xx} \over  \left [ 1+\beta^2 g^{00}g^{xx} |\vec{E}|^2 \right ]} \left [\, |\vec j|^2 + \beta^2 g^{00}  g^{xx}(\vec j \cdot \vec E )^2 \right ] \,.
\eeqa

\medskip

\subsection{The conductivity tensor}

\medskip

Following the Karch-O'Bannon's procedure, described in Section 3, we find that the reality constraint of the onshell DBI action leads to the following conditions at $z = z_\ast$ :
\beqa
Q_0 &=& 1 + \beta^2 g^{00} g^{xx} |\vec{E}|^2 + \beta^2 (g^{xx})^2 B^2 = 0 \, ,  \label{3dconstraint0} \\
\tilde j_0 &=& \beta \Big \{ - \left [ 1  + \beta^2 g^{00} g^{xx} |\vec{E}|^2 \right ] j^0
\,+\,  \beta^2 g^{00} g^{xx} \, B \vec{j} \cdot \vec{ \tilde E} \Big \} = 0 \, ,  \label{3dconstraint1}  \\
\gamma^2 \beta^4  &+& \frac{\beta^2 g^{00} g^{xx}}{\left [ 1 + \beta^2 g^{00}g^{xx} |\vec{E}|^2 \right ]} \Big[\, |\vec j|^2 + \beta^2 g^{00}  g^{xx}(\vec j \cdot \vec E )^2\Big] = 0 \label{3dconstraint2}
\end{eqnarray}
From equations (\ref{3dconstraint0}) and  (\ref{3dconstraint1}) we get \beqa
1 + \beta^2 (g^{xx})^2 B^2 &=& -  \beta^2 g^{00} g^{xx} |\vec{E}|^2 \, ,
\label{3dQ0der} \\
\vec{j} \cdot \vec{ \tilde E} &=& \frac{\beta^2 (g^{xx})^2 |\vec{E}|^2 B j^0}{\left [ 1 + \beta^2 (g^{xx})^2 B^2 \right ]} \, . \label{3djtildeEj0}
\eeqa

The current can be decomposed as
\beqa
\vec j  =  {\Big [   (\vec j \cdot \vec{\tilde E}) \vec{\tilde E}+  (\vec j \cdot \vec E )  \vec E  \big] \over |\vec{E}|^2} \label{3ddecompj} \, .
\eeqa
Using (\ref{3dQ0der}) and (\ref{3djtildeEj0}) in (\ref{3ddecompj}) we obtain
\beqa
\vec j  = - \frac{ \beta^2 g^{xx}}{ \left [ 1 + \beta^2 (g^{xx})^2 B^2 \right ] } \Big \{ g^{00} (\vec{j} \cdot \vec{E}) \vec{E}
- g^{xx} B j^0 \vec{ \tilde E} \Big \} \,. \label{3djjE}
\eeqa
Using (\ref{3dconstraint2}) and (\ref{3djjE}) we solve $\vec{j} \cdot \vec{E}$. Assuming $\vec{j} \cdot \vec{E} \ge 0$ we get
\beqa
\vec{j} \cdot \vec{E} = - \frac{g_{00}}{\beta^2}
\sqrt{(g_{xx})^2 \left [ 1 + \beta^2 (g_{xx})^2 B^2 \right ] \gamma^2 \beta^4 + \beta^2 (j^0)^2 } \,.
\eeqa

Writing the current as $j_i = \sigma_{ij} E_j $ we get the conductivity tensor
\beqa
\sigma_{ij}  &=&   { (g^{xx}) \over \left [1+ \beta^2 (g^{xx})^2 B^2 \right] } \Big \{    \sqrt{(g_{xx})^2[1+\beta^2 (g^{xx})^2 B^2] \gamma^2 \beta^4 +\, \beta^2   (j^0)^2  }  \,\delta_{ij} \cr
&+& \beta^2 g^{xx}  B j^0 \,  \epsilon_{ij} \Big \} \,. \label{3dconductivity}
\eeqa
In this case the conductivity has two identical diagonal components $\sigma_{11} = \sigma_{22}$  and two opposite non-diagonal components
$\sigma_{12} = - \sigma_{21}$. Remember that in (\ref{3dconductivity}) the parameters $\beta$ and $\alpha$ as well as the metric $g_{xx}$ are evaluated at the special point $z_\ast$ satisfying $Q_0(z_\ast)=0$.

Our result (\ref{3dconductivity}) is also consistent with that obtained recently in \cite{Kim:2011qh,Kim:2011zd}, using the method of open string metric.

\medskip

\subsection{Lifshitz symmetry and strange metals}

\medskip

Here we consider the finite temperature background considered in  \cite{Hartnoll:2009ns} in a holographic approach to strange metals. The induced metric of a Dp-brane in the massless case can be written as
\beqa
ds^2 = R^2 \Big( - {f(v)\, dt^2 \over v^{2 \bar z} } + {dv^2 \over f(v) v^2} + { dx^2 + dy^2 \over v^2} \Big) + d \Omega^2 \,, \label{Lifshitzmetric}
\eeqa
where $\Omega$ is a compact submanifold and $v$ is the radial direction that extends from the boundary $v=0$ to the horizon $v_+$.

The function $f(v)$ is model dependent but satisfies the universal condition $f(v_+)=0$. As a consequence of this condition we get
\beqa
T = \frac{|f'(v_+)|}{4 \pi v_+^{\bar z-1}} \sim \frac{1}{v_+^{\bar z}}
\eeqa

The metric (\ref{Lifshitzmetric}) realizes Lifshitz symmetry through the following scaling transformations :
\beqa
t \to \lambda^{\bar z} t \quad , \quad
\vec{x} \to \lambda \vec{x} \quad , \quad
v \to \lambda v \,.
\eeqa
The parameter $\bar z$ is then interpreted as the dynamical critical exponent of the quantum critical sector in the dual field theory.

The Dp-brane action corresponding to (\ref{Lifshitzmetric}) can be written as in (\ref{DBI2}) with
\beqa
\beta(v)  &=& 2 \pi \alpha' \, , \cr
\gamma (v)  &=& \mu_p V_\Omega e^{- \phi} =: \tau_{eff} \, , \cr
g_{xx}(v) &=& \frac{R^2}{v^2} \,.
\label{choicesLtz}
\eeqa
Therefore, from (\ref{3dconductivity}) we get for the Lifshitz background the conductivity tensor
\beqa
\sigma_{ij} &=& \frac{1}{ \left [ 1 +  ( \frac{2 \pi \alpha'}{R^2} )^2 v_\ast^4 B^2 \right ] }
\Big \{
\sqrt{ \left [ 1 +  ( \frac{2 \pi \alpha'}{R^2}  )^2 v_\ast^4 B^2 \right ] \tau_{eff}^2 (2 \pi \alpha')^4
+ ( \frac{2 \pi \alpha'}{R^2} )^2 v_\ast^4 (j^0)^2 } \, \delta_{ij} \cr
&+&  ( \frac{2 \pi \alpha'}{R^2}  )^2 v_\ast^4 B j^0  \, \epsilon_{ij} \Big \} \,.
\eeqa
This result is consistent with the one obtained in \cite{Hartnoll:2009ns}. In the limit of large densities the conductivity tensor reduces to
\beqa
\sigma_{ij} &=& \frac{( \frac{2 \pi \alpha'}{R^2} ) v_\ast^2 \, j^0 }{ \left [ 1 +  ( \frac{2 \pi \alpha'}{R^2} )^2 v_\ast^4 B^2 \right ] }
\Big \{ \delta_{ij} + ( \frac{2 \pi \alpha'}{R^2} ) v_\ast^2 B \epsilon_{ij} \Big \} \cr
&=:& \frac{ \rho^{-1}  }{ \left [ 1 +   \frac{B^2}{\mu^2 M^2} \right ] } \Big \{ \delta_{ij} + \frac{B}{\mu M} \epsilon_{ij} \Big \} \, .
\label{conductDrude2d}
\eeqa
The parameter $\rho$ is interpreted as the DC resistivity which has the following temperature dependence :
\beqa
\rho = \frac{R^2}{2 \pi \alpha' v_\ast^2 j^0}   \sim \frac{T^{2/ \bar  z}}{j^0} \,.
\eeqa
For the case $\bar z =2$ the DC resistivity is linear in the temperature, as expected for strange metals.
Note that the second line in (\ref{conductDrude2d}) is the same obtained in a $2+1$ dimensional Drude model with $\mu M$ given by
\beqa
\mu M = \frac{R^2}{2 \pi \alpha' v_\ast^2 } \sim T^{2 / \bar z} \,.
\eeqa

\section{Conclusions}

We have proposed a simple expansion method to investigate DBI equations and define holographic currents in a general class of backgrounds (whose metric depends on the radial coordinate only) and Dp-brane configurations. Our method is alternative to the traditional approach where an effective determinant is calculated for each field configuration. In our method, we write the DBI equations for  arbitrary gauge field configurations before choosing any ansatz. We expect that our results would be useful in several holographic models that are derived from String Theory. A future direction in our approach would be to extend the expansion method for the case of backgrounds with conical deficits (see for instance \cite{Bayona:2010sd}).

As an application of our method we have discussed thoroughly the situation where constant electric and magnetic fields are turned on.
  Following the Karch-O'Bannon's method we found for a general background, satisfying homogeneity and isotropy, a frame-independent expression for the conductivity tensor in $3+1$ and $2+1$ dimensions. For particular backgrounds we have recovered previous results on holographic metals and strange metals. An interesting point in that analysis was that the presence of a horizon guarantees the existence of a solution that satisfies the real constraint.

 It would be interesting to investigate also  physical systems where the electric/magnetic fields have a time or spatial dependence. We leave this problem for future work.

\section*{Acknowledgements}

The authors would like to thank the ICTP in Trieste (Italy) and the ICTP-SAIFR in Sao Paulo (Brazil) for the hospitality. The work of A.B.B. is supported by a Capes fellowship. The work of C. N. F is partially supported by a CNPq fellowship.

\vfill\eject

\appendix

\section{The Drude model}

The Drude equation for an electromagnetic current can be written as
\beqa
j^0 \vec{E} - \mu M \vec{j} + \vec{j} \times \vec{B} = 0 \, ,
\eeqa
where $\mu$ is the drag coefficient and $M$ is the mass of the charge carrier.
We can write this equation in a matrix form :
\beqa
{\cal M}_{ij} j_j = - j^0 E_i \, ,
\eeqa
where
\beqa
{\cal M}_{ij} = - \mu M \delta_{ij} + \epsilon_{ijk} B_k \,.
\eeqa
The inverse of this object is
\beqa
{\cal M}^{-1}_{ij} = - \frac{1}{\mu M \left [ 1 + \frac{|\vec{B}|^2}{\mu^2 M^2} \right ] }
\Big [ \delta_{ij} + \frac{B_i B_j}{\mu^2 M^2} + \epsilon_{ijk} \frac{B_k}{\mu M} \Big ] \,.
\eeqa
Using this result we get the Drude conductivity
\beqa
\sigma^D_{ij} = \frac{j^0}{\mu M \left [ 1 + \frac{|\vec{B}|^2}{\mu^2 M^2} \right ] }
\Big [ \delta_{ij} + \frac{B_i B_j}{\mu^2 M^2} + \epsilon_{ijk} \frac{B_k}{\mu M} \Big ] \,. \label{Drude3dv2}
\eeqa

\section{The effect of a 5-d Chern-Simons term}

In some brane models that reduce to five dimensions we obtain an effective Chern-Simons (CS) action
\begin{equation}
S_{CS} = \frac{\alpha}{4} \, \epsilon^{\ell m
  n p q} \int\! {\rm d}^4x {\rm d}z\, {\cal A}_{\ell} {\cal F}_{mn} {\cal F}_{pq} \,.
\end{equation}

In the special case of (\ref{DCansatz}), where we have constant electric and magnetic fields, the DBI-CS equations reduce to
\beqa
0 &=&  \alpha \vec{E} \cdot \vec{B}  \, , \label{DBICSAz} \\
- j^0 &=& \frac{\gamma \beta^2}{Q(z)}
\sqrt{-g} g^{zz} g^{00} \Big \{  \left [ 1 + \beta^2 (g^{xx})^2 \vec{B}^2 \right ] \partial_z f_0
 + \beta^2 (g^{xx})^2 (\vec{E} \times \vec{B}) \cdot (\partial_z \vec{f}) \Big \} \cr
&-& 6 \alpha \vec{B} \cdot \vec{f} \, , \label{j0CS} \\
- \vec{j} &=& \frac{\gamma \beta^2}{Q(z)} \sqrt{-g} g^{zz} g^{xx}
\Big \{  \left [ 1 + \beta^2 g^{00} g^{xx} |\vec{E}|^2 \right ]  \partial_z \vec{f} \cr
&-& \beta^2 g^{00} g^{xx} \left [  ( \vec{E} \cdot \partial_z \vec{f}) \vec{E}
 + (\partial_z f_0) \vec{E} \times \vec{B} \right ]
+ \beta^2 (g^{xx})^2 (\vec{B} \cdot \partial_z \vec{f}) \vec{B} \Big \} \cr
&+& 6 \alpha \left [  f_0 \vec{B} +  \vec{f} \times \vec{E} \right ] \,. \label{jiCS}
\eeqa
Here the quantities $j^0$ and $\vec{j}$ are integration constants, not necessarily the electromagnetic current\footnote{The definition of holographic currents in the presence of a Chern-Simon term is subtle because of chiral anomaly (see for instance \cite{Rebhan:2009vc,Gynther:2010ed,BallonBayona:2012wx})}.

According to eq. (\ref{DBICSAz}) the electric and magnetic
field must be orthogonal\footnote{If we want to consider parallel electric and magnetic fields we would need to introduce a
time dependence on the ${\cal A}_0$ component (see for instance \cite{Rebhan:2009vc})}. Note that we can rewrite the last term in (\ref{jiCS}) as
\beqa
6 \alpha \left [ f_0 \vec{B} +  \vec{f} \times \vec{E} \right ]  = 6 \alpha \Big \{ \left [ f_0 |\vec{B}|^2 + (\vec{E} \times \vec{B}) \cdot \vec{f} \right ] \frac{\vec{B}}{|\vec{B}|^2}
 - \frac{(\vec{B} \cdot \vec{f})}{|\vec{B}|^2} \vec{E} \times \vec{B} \Big \} \, . \label{ltjiCS}
\eeqa
Here we used the vectorial decomposition
\beqa
\vec{f} &=& \frac{1}{|\vec{E} \times \vec{B}|^2} \Big \{ \left [ (\vec{f} \cdot \vec{E}) |\vec{B}|^2 - (\vec{f} \cdot \vec{B}) (\vec{E} \cdot \vec{B}) \right ] \vec{E} \cr
&+& \left [ (\vec{f} \cdot \vec{B}) |\vec{E}|^2 - (\vec{f} \cdot \vec{E}) (\vec{E} \cdot \vec{B})  \right ] \vec{B}
+ \vec{f} \cdot (\vec{E} \times \vec{B}) \vec{E} \times \vec{B} \Big \} \, . \label{jeq1}
\eeqa
and the constraint $\vec{E} \cdot \vec{B} = 0 $.

Taking the scalar product of (\ref{jiCS}) with $\vec{E}$, $\vec{B}$ and $\vec{E} \times \vec{B}$ respectively we get
the equations :
\beqa
\sqrt{-g} g^{zz} g^{xx} \vec{E} \cdot \partial_z \vec{f} &=& - \frac{Q(z)}{\gamma \beta^2 } \vec{j} \cdot \vec{E} \, , \label{fECS} \\
\sqrt{-g} g^{zz} g^{xx} \vec{B} \cdot \partial_z \vec{f} &=&
- \frac{Q(z)}{\gamma \beta^2 Q_0(z)}  \Big \{  \vec{j} \cdot \vec{B}
 +6 \alpha \left [ f_0 |\vec{B}|^2 + (\vec{E} \times \vec{B}) \cdot \vec{f} \right ] \Big \}  \, , \label{fBCS}
\eeqa
\beqa
\sqrt{-g} g^{zz} g^{xx}  &&
\Big \{ \left [ 1 + \beta^2 g^{00} g^{xx} |\vec{E}|^2 \right ]  (\vec{E} \times \vec{B}) \cdot \partial_z \vec{f}
- \beta^2 g^{00} g^{xx} |\vec{E} \times \vec{B}|^2 \partial_z f_0 \Big \}
= \cr
&& \frac{Q(z)}{\gamma \beta^2} \Big [ \vec{j} \cdot (\vec{E} \times \vec{B})
- 6 \alpha (\vec{B} \cdot \vec{f}) |\vec{E}|^2 \Big ]
\,. \label{jEBCS}
\eeqa
Here we used $\vec{E} \cdot \vec{B} = 0$  . From (\ref{j0CS}) and (\ref{jEBCS}) we get
\beqa
\sqrt{-g} g^{zz} g^{xx} (\vec{E} \times \vec{B}) \cdot \partial_z \vec{f} &=&  - \frac{Q(z)}{\gamma \beta^2 Q_0(z) }
\Big \{ - \beta^2 (g^{xx})^2 |\vec{E} \times \vec{B}|^2 j^0 \cr
&+& \left [ 1 + \beta^2 (g^{xx})^2 |\vec{B}|^2 \right ] \vec{j} \cdot (\vec{E} \times \vec{B})
- 6 \alpha (\vec{f} \cdot \vec{B}) |\vec{E}|^2 \Big \}  \cr
&=:& - \frac{Q(z)}{\gamma \beta^5 Q_0(z)} \Big \{ \hat j_{EB} - 6 \alpha \beta^3 (\vec{f} \cdot \vec{B}) |\vec{E}|^2 \Big \}\, . \label{fEBCS}
\eeqa
\beqa
\sqrt{-g} g^{zz} g^{00} \partial_z f_0 &=& \frac{Q(z)}{\gamma \beta^2 Q_0(z) } \Big \{ - \left [ 1 + \beta^2 g^{00} g^{xx} |\vec{E}|^2 \right ] j^0 \cr
&+& \beta^2 g^{00} g^{xx} \vec{j} \cdot (\vec{E} \times \vec{B})
+ 6 \alpha \vec{f} \cdot \vec{B} \Big \}
\cr
&=:& \frac{Q(z)}{\gamma \beta^3 Q_0(z) } \Big \{ \hat j_0 + 6 \alpha \beta \vec{f} \cdot \vec{B} \Big \}  \, . \label{f0CS}
\eeqa
Using (\ref{ltjiCS}), (\ref{fECS}), (\ref{fBCS}) and (\ref{f0CS}) in (\ref{jiCS}) we get
\beqa
&& \sqrt{-g} g^{zz} g^{xx} \left [ 1 + \beta^2 g^{00} g^{xx} |\vec{E}|^2 \right ] \left [  \partial_z \vec{f} - (\vec{B} \cdot \partial_z \vec{f}) \frac{ \vec{B}}{|\vec{B}^2|} \right ] \cr
&=& \frac{Q(z)}{\gamma \beta^4 Q_0(z)} \Big \{ Q_0(z) \left [ - \beta^2 \vec{j} + \beta^2 (\vec{j} \cdot \vec{B}) \frac{ \vec{B}}{|\vec{B}|^2} \right ]
- \beta^4 Q_0(z) g^{00} g^{xx} (\vec{j} \cdot \vec{E}) \vec{E} \cr
&-& \beta^2 \Big [ (g^{xx})^2 \beta \hat j_0 - 6 \frac{\alpha}{|\vec{B}|^2} \left [ 1 + \beta^2 g^{00} g^{xx} |\vec{E}|^2 \right ] (\vec{B} \cdot \vec{f}) \Big ] \vec{E} \times \vec{B}  \Big \} \label{fialmfinalCS}
\eeqa
We can take the derivative of (\ref{fBCS}) and use (\ref{fEBCS}) and (\ref{f0CS}) to get a second order differential equation for $\vec{B} \cdot \vec{f}$ :
\beqa
&& \sqrt{-g} g^{zz} g^{xx} \left ( \frac{\gamma \beta^5 Q_0(z)}{Q(z)} \right ) \partial_z
\Big [\frac{\gamma \beta^2 Q_0(z)}{Q(z)} \sqrt{-g} g^{zz} g^{xx} \partial_z (\vec{B} \cdot  \vec{f})  \Big ]  \cr
&=& - 6 \alpha \Big \{  g_{00} g^{xx} \beta^2 |\vec{B}|^2 \hat j_0 - \hat j_{EB}
+ 6 \alpha  \left [ g_{00} g^{xx} \beta^2 |\vec{B}|^2 + \beta^2 |\vec{E}|^2 \right ] \beta \vec{B} \cdot \vec{f}\Big \} \, .
\eeqa

Finally, using (\ref{fECS}),  (\ref{fEBCS}), (\ref{f0CS})  and (\ref{fialmfinalCS}) in the definition of $Q(z)$ we get
\beqa
\frac{Q(z)}{\gamma \beta^2 Q_0(z)}  &=&  \sqrt{Q_0(z)} \left [ 1 + \beta^2 g^{zz} g^{xx} \frac{ (\vec{B} \cdot \partial_z \vec{f})^2}{|\vec{B}|^2} \right ]^{1/2}
\Big \{ \gamma^2 \beta^4 Q_0^2(z) \cr
&+&  (g^{xx})^3 \left [ 1 + \beta^2 (g^{xx})^2 |\vec{B}|^2 \right ] \left [ \hat j_0 + 6 \alpha \beta \vec{B} \cdot \vec{f} \right ]^2 \cr
&-& 2 g^{00} (g^{xx})^4 \left [ \hat j_0 + 6 \alpha \beta \vec{B} \cdot \vec{f} \right ] \left [ \hat j_{EB} - 6 \alpha \beta^3 (\vec{B} \cdot \vec{f}) |\vec{E}|^2 \right ] \cr
&-& (g^{00})^2 (g^{xx})^3 \beta^4 Q_0^2(z) (\vec{j} \cdot \vec{E})^2  \cr
&+&  \frac{ g^{00} (g^{xx})^2}{\beta^2 \left [ 1 + \beta^2 g^{zz} g^{00} |\vec{E}|^2 \right ] }
\Big | Q_0(z) \left [ - \beta^2 \vec{j} + \beta^2 (\vec{j} \cdot \vec{B})  \frac{ \vec{B}}{|\vec{B}|^2} \right ] \cr
&-& \beta^4 Q_0(z) g^{00} g^{xx} (\vec{j} \cdot \vec{E}) \vec{E} \cr
&-& \beta^2 \Big [ (g^{xx})^2 \beta \hat j_0 - 6 \frac{\alpha}{|\vec{B}|^2} \left [ 1 + \beta^2 g^{00} g^{xx} |\vec{E}|^2 \right ] (\vec{B} \cdot \vec{f}) \Big ] \vec{E} \times \vec{B}  \Big |^2
\Big \}^{-1/2} \,.
\eeqa

\end{document}